\documentstyle[aps]{revtex}
\tightenlines
\newcommand{\ycaba}{Y$_{2-x}$Ca$_x$BaNiO$_5$}

\begin{document}

\title{Holes in a Quantum Spin Liquid}
\author{Guangyong Xu$^1$, G. Aeppli$^2$, M.~E.~Bisher$^2$, C.
Broholm$^{1,3}$, 
J.~F.~DiTusa$^4$, C.~D.~Frost$^5$, T.Ito$^6$, K.~Oka$^6$, 
R.~L.~Paul$^3$, H.~Takagi$^7$, and  M.~M.~J.~Treacy$^2$}
\address{
$^1$Department of Physics and Astronomy, Johns Hopkins University,
Baltimore, Maryland 21218, USA\\
$^2$NEC Research Institute, 4 Independence Way, Princeton, New Jersey
08540, USA\\
$^3$NIST Center for Neutron Research, Gaithersburg, Maryland 20899,
USA\\
$^4$Department of Physics and Astronomy, Louisiana State University,
Baton Rouge, Louisiana 70803, USA\\
$^5$ISIS Facility, Rutherford Appleton Laboratory, Chilton, Didcot, Oxon
OX11 0QX, UK\\
$^6$Electrotechnical Laboratory, Tsukuba 305, Japan\\
$^7$Department of Advanced Materials Science,
Graduate School of Frontier Sciences,
University of Tokyo Hongo, Tokyo 113-8656, Japan}
\maketitle

{\bf
Magnetic neutron scattering provides evidence for nucleation of
antiferromagnetic droplets around impurities in a doped nickel-oxide
based quantum magnet. The undoped parent compound contains a spin
liquid with a cooperative singlet ground state and a gap in the
magnetic excitation spectrum. Calcium doping creates excitations below
the gap with an incommensurate structure factor. We show that weakly
interacting antiferromagnetic droplets with a central phase shift of
$\pi$ and a size controlled by the correlation length of the quantum
liquid can account for the data. The experiment provides a first quantitative
impression of the magnetic polarization cloud associated with holes in
a doped transition metal oxide.}

      Spin density modulations in transition metal oxides are receiving
huge attention because of possible connections to high temperature
superconductivity.  The modulations appear upon introduction of charge
carriers, via chemical substitution, into an insulating and
antiferromagnetic parent compound, and tend to be static when the
carriers are frozen, dynamic when they are mobile.  Evidence for such
modulations has been largely confined to materials whose magnetism and
charge transport are quasi-two-dimensional({\em1-4}), and whose
parent insulators are ordered antiferromagnets. We
provide evidence for analogous phenomena in a quasi-one-dimensional
oxide({\em5}), \ycaba , for which the parent is a spin liquid by virtue
of quantum fluctuations({\em6-10}), and which has motivated
considerable theoretical activity({\em11-16}).

The key features of this orthorhombic material are the chains of
NiO$_6$ octahedra (Fig.~1A). The octahedra are corner-sharing, which
results in the dominance of magnetism({\em8}) and electrical conduction
({\em 5}) by the very simple ..O-Ni-O-Ni-O.. back-bone.  The  magnetic
degree of freedom at each Ni site is the spin S=1 associated with the
3d$^8$ configuration of Ni$^{2+}$. Each of the S=1 ions is coupled to
its neighbors via antiferromagnetic (AFM) super-exchange through shared
O$^{2-}$ ions.   Replacing the off-chain Y$^{3+}$ by Ca$^{2+}$
introduces holes primarily onto apical oxygen atoms in the chains
and while the materials remain insulators, doping significantly increases 
conductivity at finite temperatures. 
Thus, \ycaba\  is a one-dimensional analog of the cuprates,
where off-$\rm CuO_2$-plane chemical impurities donate holes to the
$\rm CuO_2$ planes.

     Magnetic one-dimensionality causes the parent compound $\rm
Y_2BaNiO_5$ to be a spin liquid prevented from ordering
antiferromagnetically by quantum fluctuations. The material is not an
ordinary paramagnet with heavily damped spin fluctuations, but rather
the magnetic analog of super-fluid $^4$He because it has a
macroscopically coherent quantum ground state and lacks static order.
Affleck, Kennedy, Lieb and Tasaki (AKLT) provided a prescription for
this state({\em17,18}). They considered each S=1 degree of freedom as
the triplet ground state formed between two ferromagnetically coupled
S=1/2 degrees of freedom, in accord with atomic physics which posits a
large and positive Hunds rule coupling between the two S=1/2 holes of
Ni$^{2+}$. To understand the qualitative features of the ground and
excited states of the spin-1 chain, they started from the limit of
vanishing intra-atomic interactions between spin-1/2 holes.  Because
each site of the spin chain has as many nearest neighbors as S=1/2
degrees of freedom, there exists a macroscopic singlet with the full
symmetry of the Hamiltonian wherein all S=1/2 degrees of freedom take
part in nearest neighbor bond singlets  (Fig 2A). Excitations from this
``Valence Bond Solid'' (VBS) state are triplets formed by broken
valence bonds that propagate along the chain. The triplet band is separated 
from the VBS by an energy of order $J$ corresponding to the
energy required to create a triplet at rest.  As the Ferromagnetic (FM)
Hund's rule coupling is taken to much larger values than $J$ these low
energy states evolve adiabatically into the ground and excited states
respectively of the S=1 chain. The major change is that for energies
below the intra-atomic exchange energy, spin-1/2 degrees of freedom can
be observed only as edge states, not in the bulk.

The triplet band can be observed experimentally for $\rm Y_2BaNiO_5$
(Fig.~2A). An 
image of the magnetic neutron scattering intensity as a function of wave vector transfer
$\tilde{q}$ along the spin chains (in units of the inverse lattice
spacing) and energy transfer, $\hbar\omega$,  was obtained
using time-of-flight neutron spectroscopy at the ISIS facility in the
UK.  Magnetic neutron scattering occurs only when $\hbar\omega$ and
$\tilde{q}$ lie on a dispersion relation, thus providing  evidence for
coherent triplet propagation along spin chains.  What distinguishes
this excitation from a spin wave is that it propagates in a medium with
no static magnetic order. Key features are (i) a 9~meV Haldane
gap({\em19}), which corresponds to the threshold energy for triplet
creation, (ii) the magnetic excitation bandwidth of 62~meV that
translates into an AFM exchange constant $J=21$~meV, and (iii)
vanishing intensity for $\tilde{q}$$=n2\pi$, which is evidence for a
singlet ground state.

What happens when by Ca doping we add carriers to the quantum spin
liquid of Fig.~2A?. Our $x=0.095(5)$ sample (Fig.~2B) reveals
excitations following a similar dispersion relation as for pure $\rm
Y_2BaNiO_5$. However, near the former minimum at $\tilde{q}=\pi$, their
intensity is substantially reduced and new magnetic scattering appears
below the Haldane gap. More detailed data collected near and below the
gap (Fig.~3B) show that near the gap energy, the scattering is clearly
peaked at $\tilde{q}=\pi$ but for $\hbar\omega =3-7$~meV it is most
intense along two vertical lines, displaced symmetrically about $\pi$.
Higher statistics energy-integrated $\tilde{q}$-scans are shown (Fig.~4) that
accurately quantify the incommensurate peaks for three single crystals
with different levels of doping. Surprisingly, increased doping does
not yield a proportional increase in the shift of the peaks from
$\pi$.  Specifically, if we fit the three data sets by the sum of two
identical Lorentzians centered at $\pi\pm\delta\tilde{q}$ with
half-widths-at-half-maximum, $\tilde{\kappa}$, we find that
$\delta\tilde{q}$ extrapolates to a finite value in the dilute limit
increasing with $x$ as $\delta\tilde{q}(x)/\pi\approx 0.059(2)+0.25(3)
x$ (see inset to Fig.  4(B)). A similar $x$-dependence is found for
$\tilde{\kappa} (x)/\pi\approx 0.047(3)+0.20(6) x$ with the implication
that a single length scale, which is finite in the limit $x\rightarrow
0$, controls both $\tilde{q}(x)$ and $\tilde{\kappa}(x)$.

There are various models which could produce the incommensurate
fluctuations we have discovered in \ycaba . The first proceeds from a
charge-ordering hypothesis where the holes order periodically to
minimize Coulomb repulsion and the spin system acquires a period
commensurate with that of the hole lattice. The magnetic fluctuations
are
spin waves broadened in $\tilde{q}$ due to disorder in the hole
lattice.  Such a scenario is commonly supposed for the two-dimensional
analog, doped $\rm La_2NiO_4$, where  key experimental support came
from electron({\em20}) and neutron diffraction({\em21,22}). Thus
inspired, we used both techniques to search for hole ordering in
\ycaba\ down to 15K. The failure of the search makes the charge
ordering explanation of our experiments unlikely. Furthermore, in the
simplest picture, the hole density has the same period as the spin
density squared so that $\delta\tilde{q}(x)/\pi\approx x$, which is
entirely inconsistent with Fig. 4 (dashed line in the inset).

A second possibility is to identify the incommensurate magnetic
fluctuations with electron-hole pair excitations from the
one-dimensional
hole liquid associated with mobile S=1/2 hole propagation through the
VBS({\em11}). The wave-vectors $\pi\pm\delta\tilde{q}$  would be
vectors spanning the Fermi surface while the vertical nature of the
incommensurate streaks (Fig.~3) would imply a Fermi velocity exceeding
$0.5$~eV$\cdot{\rm\AA}$ngstrom.  The problem with this picture is that
even if we can ignore that the samples seem not to be  proper metals -
the resistivities({\em5}) are high at low T - the Fermi sea is expected
to grow in proportion to $x$ leading to $\delta\tilde{q}(x)\approx x$, a
result again contradicted by Fig. 4.

A last possibility starts from consideration({\em13}) of the magnetic
structure factor for a single static hole donated to the NiO chain by
the Ca ions. Because it carries a spin, a hole on a
super-exchange mediating oxygen atom induces an effective ferromagnetic
Ni-Ni
interaction({\em23,24}).  The hole-carrying oxygen
divides the chain into two semi-infinite segments, for which
wave-functions can be constructed from a basis set which is the direct
product of wave-functions for each segment and for the hole on the
intervening oxygen. The ground state wave-function for a semi-infinite
segment is a doublet rather than the singlet wave-function for the
unterminated chain. Pictorially, (Fig.~1B-C) 
the chain ends have unpaired spins because only one of the two AKLT
spins, which together represent the S=1 sites, can be paired with an
AKLT
spin on a neighboring site. The effective ferromagnetic interaction
between 
semi-infinite chains yields a ground state with inversion 
symmetry about the oxygen site. The edge states are not
localized on Ni$^{2+}$ atoms neighboring the hole, but extend into
the chain with AFM modulation over a distance of order the Haldane spin
correlation length, $\kappa^{-1}$. As a 
result, ground state correlations
$\langle S_0S_j\rangle$ between spins at site $0$ and site $j$, where
the FM bond occurs between sites $-1$  and $0$, should take the form
$(-1)^je^{-|j|\kappa}$ for $j\ge 0$ and $(-1)^{j+1}e^{-|j+1|\kappa}$
for $j \le -1$.

Neutron scattering measures the modulus squared Fourier transform of the
spin
polarization. For the ground state associated with a
single FM bond inserted within an infinite chain, this is 
\begin{equation}
S(\tilde{q})=\left|\frac{(1+e^\kappa)\cos(\tilde{q}/2)}{\cosh{\kappa}+\cos\tilde{q}}\right|^2.
\label{eqn1}
\end{equation}
Eq.~1 describes a function with nodes for
$\tilde{q}=(2n+1)\pi$ and 
pairs
of peaks, with half widths and incommensurability of order $\kappa$ 
(see dashed line in Fig.~4A).  Thus,
the description leading to Eq.~1 does have a chance of accounting for
a key feature of our data.  There are however three 
difficulties, the first being that the nodes predicted by Eq.~1 are
not
seen experimentally. The second problem is that we have neglected the
spin on the intervening oxygen. The third is that Eq.~1
describes the ground state for a single FM bond, implying that neutron
scattering should be elastic, not inelastic as seen
experimentally.  

The first difficulty has a simple resolution. At
finite hole densities the
polarization clouds overlap and the isolated impurity model is
inadequate.
A crude model, which considers the overlaps, simply truncates the
polarization clouds at neighboring impurity sites. Because FM impurity
bonds
are randomly distributed, the inversion symmetry characterizing
isolated impurities is broken thus allowing intensity at
$\tilde{q}=\pi$. 
Because we know the impurity density, $x$, from neutron activation
analysis, 
the only parameters in such a description are the extent of the
polarization
cloud, $\kappa^{-1}$, which we adjusted to
optimize the fit to our data. As shown
by the red lines in Fig.~4 the model provides a good account of the
data with $\kappa^{-1}=8.1(2), 7.3(2)$, and $7.2(5)$ for $x=0.04,
0.095$, and $0.14$
respectively. These
values are close to the exponential decay length of 6.03
calculated for the AFM spin polarization at the end of an S=1
chain({\em25}).

The modeling described so far does not include the spins of the holes
responsible for the effective FM couplings between Ni$^{2+}$ ions. The
holes reside in oxygen orbitals of \ycaba ({\em5}), but are almost
certainly not confined to single, isolated oxygens. We have consequently
generalized Eq.~1 to take into account the hole spins, with - for the
sake of definiteness - the same net amplitude as either of the Ni$^{2+}$
spins next to the FM bond, and distributed (with exponential decay) over
$\ell$ lattice sites centered on the FM bond. As long as $\ell$ exceeds
the modest value of 2, comparable to the localization length deduced
from transport data ({\em5}), the pronounced asymmetry about
$\pi$ which occurs when $\ell=0$ is relieved sufficiently to produce
fits indistinguishable from those in Fig. 4. 

How do we account for the inelasticity of the incommensurate signal?
One approach is to view the chain as consisting not of the original S=1
degrees of freedom but of the composite spin
degrees of freedom induced around holes.  The
latter interact via overlapping AFM
polarization clouds and hole wave functions, to produce effective
couplings of random
sign because the impurity spacing can be even
or odd multiples of the Ni-Ni separation. With weak interchain coupling
the ground state is likely to be a spin glass, as deduced
from other experiments({\em26}) on \ycaba . The
``incommensurate'' nature of the excitations continues to follow from
the
structure factor of the spin part of the hole wavefunctions.

      In summary, we have measured the
magnetic fluctuations in single crystals of a doped one-dimensional spin
liquid. At
energies above
the spin gap, the triplet excitations of the parent compound, $\rm
Y_2BaNiO_5$, persist
with doping. However, below
the gap, we find new excitations with a broad spectrum and
characteristic wave vectors that are displaced from the zone
boundary by an amount of order the inverse correlation-length for the
parent. The incommensurate fluctuations, encountered here in a doped
one-dimensional transition metal oxide, resemble those seen in metallic
cuprates. However,  
one-dimensionality makes them
easier to model for $\rm Y_2BaNiO_5$,  and our analysis reveals that
``incommensurate'' peaks
arise naturally even without hole order because of the characteristic
spin density modulation that develops around a defect in the singlet
ground state of a quantum spin liquid. This phenomenon accounts for the
weak dependence of the incommensurability on doping in the
one-dimensional nickelate. Indeed, \ycaba\ 
gives us our first quantitative impression of the magnetic polarization
cloud associated with the holes in a doped transition metal oxide. Our
results imply that the spin part of the hole wave-function is actually
the edge state nucleated by the hole in a quantum spin fluid.

\newpage
\begin{center}{\Large References}
\end{center}
  
\begin{enumerate}
\item  S.-W. Cheong et al.,  {\it Phys. Rev. Lett.} {\bf 67}, 1791
(1991). 
\item  S. M. Hayden et al., {\it Phys. Rev. Lett.} {\bf 68}, 1061
(1992).
\item  H. A. Mook et al., {\it Nature} {\bf 395}, 580 (1998).
\item  J. M Tranquada, B. J. Sternlieb, J. D. Axe, Y. Nakamura, S.
Uchida, {\it Nature} {\bf 375}, 561 (1995).
\item  J. F. DiTusa et al., {\it Phys. Rev. Lett.} {\bf 73}, 1857
(1994).
\item  J. Darriet and L.P. Regnault, {\it Solid State Commun.}
{\bf 86}, 409 (1993).
\item  J. F. DiTusa et al., {\it Physica B} {\bf 194-196}, 181
(1994).
\item  G. Xu, et al., {\it Phys. Rev. B} {\bf 54}, R6827 (1996).
\bibitem{9}  T. Sakaguchi, K. Kakurai, T. Yokoo,  J. Akimitsu, {\it
Phys. Soc. Japan}  {\bf 65}, 3025 (1996).
\item A. Keren et al., {\it Phys. Rev. Lett.} {\bf 74},
3471 (1995).
\item  E. Dagotto, J. Riera, A. Sandvik, A. Moreo, {\it Phys.
Rev. Lett.} {\bf 76}, 1731 (1996).
\item  Z.-Y. Lu, Z.-B. Su, L. Yu, {\it Phys. Rev. Lett.} {\bf
74}, 4297 (1995).
\item  K. Penc and H. Shiba, {\it Phys. Rev. B} {\bf 52}, R715
(1995).
\item  S. Fujimoto and N. Kawakami, {\it Phys. Rev. B} {\bf 52},
6189 (1995).
\item  E. S. Sorensen and I. Affleck, {\it Phys.Rev. B} {\bf 51},
16115 (1995).
\item  R. A. Hyman, K. Yang, R. N. Bhatt, S. M. Girvin, {\it
Phys. Rev. Lett.} {\bf 76}, 839 (1996).
\item  I. Affleck, T. Kennedy, E. H. Lieb, H. Tasaki, {\it Phys.
Rev. Lett.} 
{\bf 59}, 799 (1987). 
\item  T. Kennedy, {\it J.  Phys. Cond. Matter} {\bf 2}, 5737
(1990).
\item  F. D. M. Haldane, {\it Phys. Lett.}  {\bf 93A}, 464
(1983).
\item  C. H. Chen, S.-W. Cheong, A. S. Cooper, {\it Phys. Rev.
Lett.} {\bf 71}, 2461 (1993).
\item  J. M. Tranquada, D. J. Buttrey, V. Sachan, J. E. Lorenzo,
{\it Phys. Rev.  Lett.} {\bf 73}, 1003 (1994).
\item  S.-H. Lee and S.-W. Cheong, {\it Phys. Rev. Lett.} {\bf
79}, 2514 (1997)
\item V.J.Emery and G. Reiter, {\it Phys. Rev. B} {\bf 38},
4547(1988).
\item A. Aharony et al., {\it Phys. Rev. Lett.} {\bf 60},
1330(1988).
\item  S. R. White and D. A. Huse, {\it Phys. Rev. B} {\bf 48},
3844 (1993).
\item  K. Kojima, et al., {\it Phys. Rev. Lett.} {\bf 74}, 3471
(1995).
\item
We thank T. M. Rice, A. J. Millis, and Q. Huang for useful discussions
and
A. Krishnan for TEM measurements. Work at JHU was supported by
the NSF under DMR-9453362. Work at LSU was supported by the NSF
under DMR-9702690 and the State of Louisiana Board of Regents under
LEQSF(RF/1996-99)-RD-A-05. This work utilized neutron research
facilities supported by NIST and the NSF through DMR-9423101.

\end{enumerate}

\newpage

\begin{figure}
\caption{(A)~Chain unit of $\rm Y_2BaNiO_5$ 
featuring nickel atoms with octahedral oxygen coordination and Ca$^{2+}$
impurities on Y$^{3+}$ sites. (B)~Schematic of S=1/2 degrees of
freedom 
at the edges of the AKLT state. (C)~Ferromagnetically coupled chain-end
S=1/2 degrees of freedom.}
\label{figure1}
\end{figure}

\begin{figure}
\caption{Overview of magnetic fluctuations in (A) 
pure and (B) $9.5\%$ doped $\rm Y_2BaNiO_5$ at T=10K. The 
initial neutron beam energy was 90~meV and the 
chain axis was perpendicular to the incident beam direction. Boxes
indicate
regions examined in Fig.~3. The color
bar shows values for $\frac{ki}{kf}\frac{d^2\sigma}{d\Omega dE_f}$ in
units of mbarn meV$^{-1}$ per Ni.}
\label{figure2}
\end{figure}

\begin{figure}

\caption{Low energy detail of magnetic excitations in (A) 
pure and (B) $9.5\%$ doped $\rm Y_2BaNiO_5$. (A) shows
time-of-flight data  at
T=10K (MARI spectrometer, ISIS pulsed neutron source) while
(B) shows data collected at T=1.5K using triple-axis spectrometers (SPINS
and BT2 at NIST steady state nuclear reactor with final energies 5~meV
and 14.7~meV respectively). The color
bar shows values for $\frac{ki}{kf}\frac{d^2\sigma}{d\Omega dE_f}$ in
units of mbarn meV$^{-1}$ per Ni.}
\label{figure3}
\end{figure}

\begin{figure}
\caption{$\tilde{q}$-scans, collected using SPINS at NIST, through the
incommensurate peaks
at fixed energy transfer of 4.5~meV for three different hole
concentrations.
The energy resolution of the spectrometer was 2~meV FWHM. The dashed
green line
in A shows the single impurity model Eq.~1 convolved with the
instrumental resolution (solid bar). The red lines
take into account that neighboring impurities truncate the spin
polarization
cloud around an impurity bond. Inset in B shows half the distance
$\delta\tilde{q}(x)/\pi$ between the peaks of two Lorentzians superposed
to fit the data. }
\label{figure4}
\end{figure}

\end{document}